\documentclass[12pt,twoside]{article}
\usepackage{amssymb}
\usepackage{graphicx}
\usepackage{hyperref} 
\usepackage{cite}  
\begin{document}  
\begin{center}
{\large\bf On the deformed oscillator and the deformed derivative associated with the Tsallis $q$-exponential} \\ 

\bigskip

Ramaswamy Jagannathan$^{*}$\footnote{Retired Faculty (Physics); {\em email}: jagan@imsc.res.in \\ 
ORCID ID: \url{https://orcid.org/0000-0003-2968-2044}} 
and 
Sameen Ahmed Khan$^{\$}$\footnote{{\em email}: rohelakhan@yahoo.com \\ 
ORCID ID: \url{https://orcid.org/0000-0003-1264-2302}} \\ 

\medskip 

$^{*}${\it The Institute of Mathematical Sciences \\ 
4th Cross Street, Central Institutes of Technology} ({\it CIT}) {\it Campus \\  
Tharamani, Chennai 600113, India}  \\  
$^{\$}${\it Department of Mathematics and Sciences \\ 
College of Arts and Applied Sciences} ({\it CAAS}), {\it Dhofar University \\ 
Post Box No. 2509, Postal Code: 211, Salalah, Oman}  
\end{center}  

\begin{abstract}
The Tsallis $q$-exponential function $e_q(x) = (1+(1-q)x)^{\frac{1}{1-q}}$ is found to be associated with the deformed oscillator 
defined by the relations $\left[N,a^\dagger\right] = a^\dagger$, $[N,a] = -a$, and $\left[a,a^\dagger\right] = \phi_T(N+1)-\phi_T(N)$, 
with $\phi_T(N) = N/(1+(q-1)(N-1))$.  In a Bargmann-like representation of this deformed oscillator the annihilation operator $a$ 
corresponds to a deformed derivative with the Tsallis $q$-exponential functions as its eigenfunctions, and the Tsallis $q$-exponential 
functions become the coherent states of the deformed oscillator.  When $q = 2$ these deformed oscillator coherent states correspond 
to states known variously as phase coherent states, harmonious states, or pseudothermal states.  Further, when $q = 1$ this deformed 
oscillator is a canonical boson oscillator, when $1 < q < 2$ its ground state energy is same as for a boson and the excited energy levels 
lie in a band of finite width, and when $q \longrightarrow 2$ it becomes a two-level system with a nondegenerate ground state and an 
infinitely degenerate excited state.  
\end{abstract}  

\medskip 


\section{Introduction} 
{\em A posteriori} knowledge gained from special relativity and quantum mechanics suggests that it may be possible to build models in 
physics actually realizable in Nature by mathematically deforming  existing theories (see, {\em e.g.}, \cite{Sternheimer2006} and references 
therein).  We may look at the attempts to build $q$-deformed models in physics using $q$-deformations of the mathematical structures 
associated with the existing models from this point of view.  There have been two totally different $q$-deformation schemes in physics 
running in  parallel since 1980s: One emerged, essentially, from studies on quantum algebras (see, {\em e.g.}, \cite{Curtright1991} and 
references therein) and the other stemmed from the pioneering paper of Tsallis \cite{Tsallis1988} on nonextensive statistical mechanics.     

Studies on the deformation of quantum commutation relations started in 1970s from different points of view (see \cite{Arik1976, Kuryshkin1980, 
Jannussis1981}; see \cite{Mansour2016} for an account of the early history of deformed quantum commutation relations).  With the advent of 
quantum algebras in 1980s, representation theory of quantum algebras led to the $q$-deformed quantum harmonic oscillator, or the so 
called $q$-oscillator \cite{Macfarlane1989, Biedenharn1989, Sun1989, Hayashi1990}.  Harmonic oscillator being a basic paradigm playing a 
central role in modeling various physical phenomena $q$-oscillators and their generalizations have been studied extensively.  Deformed oscillators 
and related mathematical structures have been used in building models in several areas of physics: nuclear structure physics using models like 
shell model, interacting boson model, etc., vibrational-rotational molecular spectroscopy, statistics of deformed oscillators, statistics interpolating between the Bose statistics and the Fermi statistics, deformed thermodynamics and applications to condensed matter physics, nonclassical states 
in quantum optics, noncommutative probability theory, information theory, etc. (see, {\em e.g.}, \cite{Bonatsos1999, Raychev1995, Chaichian1996, Chaichian1993, Shanta1994, Sunilkumar2000, Lavagno2002, Swamy2006, Chung2019a, Marinho2012, Tristant2014, Haghshenasfard2013, 
Dey2015, Dey2016, Blitvic2012, Wada2007}, and references therein).  In the discussion of vibrational-rotational spectroscopy, besides the 
$q$-deformation of the quantum harmonic  oscillator,  the quantum rigid rotator is $q$-deformed based on the $q$-deformation of the Lie 
algebra of angular momentum operators.  

The other $q$-deformation scheme follows from the nonextensive entropy 
\begin{equation}
S_q = k\left(\frac{1-\sum_{i=1}^W p_i^q}{q-1}\right), \quad \mbox{with}\ \sum_{i=1}^W p_i = 1, 
\end{equation} 
introduced by Tsallis in \cite{Tsallis1988} where $W$ is the total number of possible configurations of the system under consideration, $p_i$ is 
the probability that the system is in the $i$th configuration, and $q$ is a real parameter.  Defining 
\begin{equation}
\ln_q x = \frac{x^{1-q}-1}{1-q}, 	
\label{qln} 
\end{equation} 
the Tsallis $q$-logarithm, we get 
\begin{equation}
S_q = k\sum_{i=1}^W p_i\ln_q\left(p_i^{-1}\right), 
\end{equation} 
such that 
\begin{equation}
\lim_{q\longrightarrow 1}S_q = -k\sum_{i=1}^W p_i\ln p_i, 
\end{equation} 
the Boltzmann-Gibbs-Shannon entropy.  From (\ref{qln}) we have   
\begin{equation}
x = \left(1+(1-q)\ln_q x\right)^{\frac{1}{1-q}}.  
\end{equation} 
Thus, if we define 
\begin{equation}
e_q(x) = (1+(1-q)x)^{\frac{1}{1-q}},    
\label{Tsallis} 
\end{equation} 
the Tsallis $q$-exponential function, then 
\begin{equation}
e_q(\ln_q x) = x.  
\end{equation} 
Note that 
\begin{equation} 
\lim_{q\longrightarrow 1}\ln_q x = \ln x, \quad \lim_{q\longrightarrow 1}e_q(x) = e^x.  
\end{equation} 
It is to be observed that in the definition of the ordinary exponential function,  
\begin{equation}
e^x = \lim_{N\longrightarrow\infty}\left(1+\frac{x}{N}\right)^N, 
\end{equation} 
if we replace the inifinitesimally small $1/N$ corresponding to large $N$ by $1-q$, with $0 < 1-q = \epsilon \approx 0$, then we get the 
Tsallis $q$-exponential function (\ref{Tsallis}).  Having understood $e_q(x)$ in this way its definition can be extended to other values of 
$q$, $<1$ or $>1$. In general, for $q \neq 1$ the definition is 
\begin{equation}
e_q(x) = \left\{\begin{array}{l} 
                (1+(1-q)x)^{\frac{1}{1-q}}\ \ \mbox{if}\ 1+(1-q)x \geq 0, \\ 
                0\ \ \mbox{otherwise}.   
                \end{array}\right. 
\label{eqxdef}
\end{equation}
It follows that $e_q(0) = 1$, $0 \leq e_q(x) \leq \infty$, and 
\begin{equation} 
\frac{de_q(x)}{dx} = \left(e_q(x)\right)^q.  
\end{equation}
It is not surprising that being a natural deformation of $e^x$ the Tsallis $q$-exponential function $e_q(x)$ and its inverse function $\ln_q(x)$ 
find applications in several areas of physics and other sciences like nonequilibrium processes, anomalous diffusion, turbulence, spin-glasses, 
nonlinear dynamics, high energy particle collisions, dissipative optical lattices, atmospheric physics, astrophysics, biological systems, medical 
imaging, earthquake studies, neural networks, economics, stock markets, etc. (see, {\em e.g.}, \cite{Gell-Mann2004, Tsallis2009, Kowalski2013, Tsallis2019, Naudts2011} and references therein, and the extensive bibliography in \cite{TsallisBiblio}).    

It should be noted that in nonextensive statistical mechanics and its applications one defines 
\begin{equation} 
e_q(kx) = (1+(1-q)kx)^{\frac{1}{1-q}}, 
\end{equation} 
and $e_q(kx) \neq e_q(x)^k$.  For example, to define the $q$-Gaussian function one takes 
$e_q\left(-\beta x^2\right) = \left(1-(1-q)\beta x^2\right)^{1/(1-q)}$.   Carlitz introduced a deformed analogue of the exponential while defining 
the socalled degenerate Bernoulli numbers and polynomials (see \cite{Carlitz1956, Carlitz1979}).   The deformed exponential function of Carlitz 
is defined by $\exp_\lambda(x) = (1+\lambda x)^{1/\lambda}$ and 
$\exp_\lambda(tx) = \left(\exp_\lambda(x)\right)^t = (1+\lambda x)^{t/\lambda}$ .   This difference between the deformed exponential of 
Carlitz and the Tsallis $q$-exponential is crucial in applications (see, {\em e.g.}, \cite{Balamurugan2016}).  

In this paper we are concerned only with the mathematical structures of $q$-deformations of the quantum harmonic oscillator.  The canonical harmonic oscillator, or the boson, is defined by the commutation relations  
\begin{equation} 
\left[N,b^\dagger\right] = b^\dagger,  \quad  \left[N,b\right] = -b,  \quad  \left[b,b^\dagger\right] = 1,   
\label{bosealg}
\end{equation} 
where $b^\dagger$, $b$ and $N$ are the boson creation, annihilation and number operators, respectively.  These relations imply that 
$b^\dagger b = N$ and $bb^\dagger = N+1$.  The Fock representation of the boson algebra (\ref{bosealg}) is given by  
\begin{equation}
b^\dagger|n\rangle = \sqrt{n+1}\,|n+1\rangle, \quad b|n\rangle = \sqrt{n}\,|n-1\rangle, \quad 
N|n\rangle = n|n\rangle, \quad n = 0,1,2,\ldots\ , 
\label{Fockrep}
\end{equation}  
with $\{|n\rangle |n=0,1,2,\ldots\}$ being an orthonormal basis.  Starting with the ground state $|0\rangle$ we can build the higher states as  
\begin{equation}
|n\rangle = \frac{{b^\dagger}^n}{\sqrt{n!}}\,|0\rangle.  
\end{equation} 
The Hamiltonian of the boson oscillator is 
\begin{equation} 
H = \left(b^\dagger b+\frac{1}{2}\right) = \frac{1}{2}\left(bb^\dagger+b^\dagger b\right), 
\label{bosonH}
\end{equation} 
with the energy spectrum 
\begin{equation}
E_n = n+\frac{1}{2}, \qquad n = 0,1,2,\ldots\ , 
\label{bosonE}
\end{equation} 
such that 
\begin{equation}
H|n\rangle = E_n|n\rangle.  
\end{equation}

Coherent states are eigenfunctions of $b$: 
\begin{equation}
b|\alpha\rangle = \alpha|\alpha\rangle, 
\end{equation} 
where $\alpha$ is any complex number.  The solution for $|\alpha\rangle$ is
\begin{equation}
|\alpha\rangle 
   = \frac{1}{\sqrt{e^{|\alpha|^2}}}\,\sum_{n=0}^{\infty} \frac{\alpha^n}{\sqrt{n!}}\,|n\rangle, 
\end{equation} 
with the normalization $\langle\alpha|\alpha\rangle = 1$.  We can write  
\begin{equation} 
|\alpha\rangle = \frac{1}{\sqrt{e^{|\alpha|^2}}}\,e^{\alpha b^\dagger}|0\rangle.  
\end{equation} 
In the Bargmann-like representation  
\begin{equation} 
b^\dagger = x, \quad b = D = \frac{d}{dx}, \quad N = xD,   
\end{equation} 
and the monomials 
\begin{equation}
\xi_n(x) = \frac{x^n}{\sqrt{n!}}, \qquad n = 0,1,2,\ldots\ , 
\end{equation} 
carry the Fock representation (\ref{Fockrep}) as given by    
\begin{equation}
b^\dagger\xi_n = \sqrt{n+1}\,\xi_{n+1},  \quad  b\xi_n = \sqrt{n}\,\xi_{n-1},  \quad 
N\xi_n  = n\xi_n, \qquad n = 0,1,2,\ldots\ . 
\end{equation} 
The monomials $\left\{\xi_n(x)\right.\left|n = 0,1,2,\ldots \right\}$ are seen to form an orthonormal basis with respect to the inner product 
\begin{equation}
\langle f|g\rangle = \left. \left[f^*(D)g(x)\right]\right|_{x=0}.  
\label{innerproduct}
\end{equation}
Note that in this representation $e^{\alpha x}$ is an eigenfunction of $b$ for any complex number $\alpha$.  Under the inner product (\ref{innerproduct}) it is seen that $\left\langle e^{\alpha x}\right.\left|e^{\alpha x}\right\rangle = e^{|\alpha|^2}$.  Thus, in this 
representation, the coherent states are given by 
\begin{equation} 
\psi_\alpha(x) = \frac{1}{\sqrt{e^{|\alpha|^2}}}\,e^{\alpha x} 
                    = \frac{1}{\sqrt{e^{|\alpha|^2}}}\,\sum_{n=0}^{\infty} \frac{\alpha^n}{\sqrt{n!}}\,\xi_n(x),   
\end{equation} 
which are normalized as $\left\langle\psi_\alpha\right.\left|\psi_\alpha\right\rangle = 1$.  

In general, a deformed oscillator algebra is prescribed by the relations 
\begin{eqnarray}
\left[N,a^\dagger\right] & = & a^\dagger,  \quad  \left[N,a\right] = -a,  \nonumber \\ 
\left[a,a^\dagger\right] & = & \phi(N+1)-\phi(N), \ \ \mbox{or}\ \  aa^\dagger = \phi(N+1),\ \ a^\dagger a = \phi(N),                                 
\label{phialg}
\end{eqnarray} 
where $a^\dagger$, $a$, and $N$ are the deformed oscillator creation, annihilation, and number operators, respectively.  The real function 
$\phi(n)$, sometimes called the deformation, or structure,  function, characterizes the deformed oscillator.  Note that the first two relations of (\ref{phialg}) imply that $a^\dagger a$ and $aa^\dagger$ commute with $N$.  In the case of the canonical boson oscillator (\ref{bosealg}) 
$\phi(N) = N$.  

The Fock representation of the algebra (\ref{phialg}) can be constructed easily as follows: 
\begin{eqnarray}
a^\dagger|n\rangle & = & \sqrt{\phi(n+1)}\,|n+1\rangle,  \quad  a|n\rangle = \sqrt{\phi(n)}\,|n-1\rangle,  \nonumber \\ 
N|n\rangle & = & n|n\rangle,  \quad  n = 0,1,2,\ldots\ , 
\label{deformedFockrep}
\end{eqnarray} 
Note that $a^\dagger a|n\rangle = \phi(n)|n\rangle$.  Since $a|0\rangle = 0$, we must have 
$\phi(0) = 0$.  Now, defining 
\begin{equation}
\phi(0)! = 1,  \quad  \phi(n)! =  \prod_{j=1}^n \phi(j),  \quad  n = 1,2,\ldots\ , 
\end{equation} 
we can write 
\begin{equation} 
|n\rangle = \frac{{a^\dagger}^n}{\sqrt{\phi(n)!}}\,|0\rangle.
\end{equation} 
The Hamiltonian of the deformed oscillator is taken to be 
\begin{equation} 
H_\phi = \frac{1}{2}\left(aa^\dagger + a^\dagger a\right), 
\end{equation} 
generalizing (\ref{bosonH}).  The energy spectrum of the deformed oscillator becomes 
\begin{equation}
E_{n,\phi} = \frac{1}{2}(\phi(n+1)+\phi(n)),  
\label{defoscE}
\end{equation}  
generalizing (\ref{bosonE}), and  
\begin{equation}
H_\phi|n\rangle = E_{n,\phi}|n\rangle.  
\end{equation} 

Associated to any $\phi(n)$ let us define the deformed exponential function 
\begin{equation}
e_\phi^x = \sum_{n=0}^{\infty} \frac{x^n}{\phi(n)!}, 
\label{ephix}
\end{equation} 
which we shall refer to as the $\phi$-exponential function.  It can be verified that the normalized coherent states of the deformed oscillator (\ref{phialg}), eigenstates of $a$, are given by 
\begin{equation}
|\alpha\rangle_{\phi} = \frac{1}{\sqrt{e_\phi^{|\alpha|^2}}}\,\sum_{n=0}^{\infty} \frac{\alpha^n}{\sqrt{\phi(n)!}}\,|n\rangle 
                               = \frac{1}{\sqrt{e_\phi^{|\alpha|^2}}}\,e_\phi^{\alpha a^\dagger}|0\rangle,    
\end{equation} 
in which the complex number $\alpha$ is such that $e_\phi^{|\alpha|^2} < \infty$.  Let 
\begin{equation} 
D_\phi F(x) = \left[\frac{1}{x}\phi(xD)\right]F(x). 
\end{equation} 
Note that as a generalization of the relation  
\begin{equation} 
Dx^n = nx^{n-1}.  
\end{equation}
we have 
\begin{equation}
D_\phi x^n = \phi(n)x^{n-1}.   
\end{equation} 
The Bargmann-like representation is 
\begin{equation} 
a^\dagger = x, \quad a = D_\phi, \quad N = xD.  
\label{phiBargmann}
\end{equation} 
so that the monomials 
\begin{equation}
\xi_{n,\phi}(x) = \frac{x^n}{\sqrt{\phi(n)!}}, \qquad n = 0,1,2,\ldots\ , 
\end{equation} 
carry the Fock representation (\ref{deformedFockrep}) as given by    
\begin{eqnarray}
a^\dagger\xi_{n,\phi} & = & \sqrt{\phi(n+1)}\,\xi_{n+1,\phi},  \quad  a\xi_{n,\phi} = \sqrt{\phi(n)}\,\xi_{n-1,\phi},  \nonumber \\  
N\xi_{n,\phi} & = & n\xi_{n,\phi},  \quad  n = 0,1,2,\ldots\ . 
\end{eqnarray} 
The monomials $\{\xi_{n,\phi}(x)|n = 0,1,2,\ldots \}$ form an orthonormal basis with respect to the inner product 
\begin{equation}
\langle f|g\rangle_\phi = \left. \left[f^*\left(D_\phi\right)g(x)\right]\right|_{x=0}.  
\label{phiinnerproduct}
\end{equation}
Note that in this representation $e_\phi^{\alpha x}$ is an eigenfunction of $a$ for the complex number $\alpha$.  Under the inner product (\ref{phiinnerproduct}) 
\begin{equation} 
\left\langle e_\phi^{\alpha x}\right.\left|e_\phi^{\alpha x}\right\rangle = e_\phi^{|\alpha|^2}. 
\end{equation} 
Thus, in this representation, the coherent states are given by 
\begin{equation} 
\psi_{\alpha,\phi}(x) = \frac{1}{\sqrt{e_\phi^{|\alpha|^2}}}\,e_\phi^{\alpha x} 
   = \frac{1}{\sqrt{e_\phi^{|\alpha|^2}}}\,\sum_{n=0}^{\infty}\frac{\alpha^n}{\sqrt{\phi(n)!}}\,\xi_{n,\phi}(x),      
\end{equation} 
with the normalization $\left\langle\psi_{\alpha,\phi}\right.\left|\psi_{\alpha,\phi}\right\rangle = 1$, for any $\alpha$ such that 
$e_\phi^{|\alpha|^2} < \infty$.  This leads to the result that the eigenfunctions of the deformed derivative $D_\phi$, or $a$, are the 
deformed exponential functions $e_\phi^{\alpha x}$ representing, apart from the normalization factors, the corresponding coherent states of 
the deformed oscillator in the Bargmann-like representation.  The deformed derivative $D_\phi$ has been used to develop a very general theory 
of deformation of classical hypergeometric functions \cite{Gelfand1998}.  

Now, an interesting question is whether there exists a deformed oscillator with which the Tsallis $q$-exponential function (\ref{Tsallis}) is 
associated such that corresponding coherent states are given by 
$\left\{e_q(\alpha a^\dagger)|0\rangle \left.\right|\alpha \in \mathbb{C}\right\}$ apart from the normalization factors.  Or, in other words, is 
there a deformed oscillator for which in the Bargmann-like representation the annihilation operator $a$ will be a deformed derivative with the 
Tsallis $q$-exponential functions as its eigenfunctions.  In the following we present the deformed oscillator and the deformed derivative associated 
with the Tsallis $q$-exponential function, after reviewing briefly some examples of deformed oscillators.  

\section{Deformed oscillators: Some examples} 
The $q$-oscillator with 
\begin{eqnarray}
\left[N,a^\dagger\right] & = & a^\dagger,  \quad  \left[N,a\right] = -a,  \quad  \left[a,a^\dagger\right] = \phi(N+1)-\phi(N),  \nonumber \\ 
\phi(N) & = & \frac{1-q^N}{1-q} = [N]_q,    
\label{qoscalg} 
\end{eqnarray} 
was studied much before the advent of quantum groups (see \cite{Arik1976, Kuryshkin1980, Jannussis1981, Mansour2016}).  Since 
\begin{equation}
[N+1]_q-q[N]_q = 1, 
\end{equation} 
the relation between $a$ and $a^\dagger$ is written as  
\begin{equation} 
aa^\dagger - qa^\dagger a = 1.      
\end{equation} 
The Fock representation is 
\begin{eqnarray}
a^\dagger|n\rangle & = & \sqrt{[n+1]_q}\,|n+1\rangle,  \quad  a|n\rangle = \sqrt{[n]_q}\,|n-1\rangle,  \nonumber \\  
N|n\rangle & = & n|n\rangle,  \quad  n = 0,1,2,\ldots\ ,  
\label{qFockrep}
\end{eqnarray} 
where 
\begin{equation} 
[n]_q = \frac{1-q^n}{1-q} 
\label{qno}
\end{equation} 
is Heine's basic number, or $q$-number (see, {\em e.g.}, \cite{Exton1983, Kac2002}).  Note that 
\begin{equation} 
[0]_q = 0, \quad  [1]_q = 1,  \quad  \lim_{q\longrightarrow 1}[n]_q = n.   
\end{equation} 
Now, with the definitions 
\begin{equation}
[0]_q! = 1,  \quad  [n]_q! = \prod_{j=1}^n [j]_q,  \quad  n = 1,2,\ldots\ , 
\end{equation} 
we have 
\begin{equation} 
|n\rangle = \frac{{a^\dagger}^n}{\sqrt{[n]_q!}}\,|0\rangle.
\end{equation}
Defining the original $q$-exponential function 
\begin{equation}
e_q^x = \sum_{n=0}^{\infty} \frac{x^n}{[n]_q!}, 
\label{qexp}
\end{equation} 
the normalized coherent states of the $q$-oscillator (\ref{qoscalg}) are given by 
\begin{equation}
|\alpha\rangle_q = \frac{1}{\sqrt{e_q^{|\alpha|^2}}}\,\sum_{n=0}^{\infty} \frac{\alpha^n}{\sqrt{[n]_q!}}\,|n\rangle 
   = \frac{1}{\sqrt{e_q^{|\alpha|^2}}}\,e_q^{\alpha a^\dagger}|0\rangle.  
\end{equation}
In the Bargmann-like representation  
\begin{equation} 
a^\dagger = x, \quad 
a = D_q = \frac{1}{x}[xD]_q = \frac{1}{x}\left(\frac{1-q^{xD}}{1-q}\right),  \quad  N  = xD,  
\end{equation} 
where $D_q$ is the Jackson derivative, or $q$-derivative, operator (see, {\em e.g.}, \cite{Exton1983, Kac2002}) such that 
\begin{equation}
D_qF(x) = \frac{F(x)-F(qx)}{(1-q)x}.  
\label{qD}
\end{equation} 
Note that 
\begin{equation} 
D_qx^n = [n]_qx^{n-1}. 
\label{Dqxn} 
\end{equation} 
In the Bargmann-like representation the monomials 
\begin{equation}
\xi_{n,q}(x) = \frac{x^n}{\sqrt{[n]_q!}}, \quad  n = 0,1,2,\ldots\ ,  
\end{equation}  
carry the Fock representation of the $q$-oscillator algebra (\ref{qoscalg}) as given by 
\begin{eqnarray}
a^\dagger\xi_{n,q} & = & \sqrt{[n+1]_q}\,\xi_{n+1,q},  \quad  a\xi_{n,q} = \sqrt{[n]_q}\,\xi_{n-1,q},  \nonumber \\  
N\xi_{n,q} & = & n\xi_{n,q},  \quad  n = 0,1,2,\ldots\ . 
\end{eqnarray} 
The monomials $\{\xi_{n,q}(x)|n = 0,1,2,\ldots \}$ form an orthonormal basis with respect to the inner product 
\begin{equation}
\langle f|g\rangle_q = \left. \left[f^*\left(D_q\right)g(x)\right]\right|_{x=0}.  
\end{equation}
It follows from (\ref{Dqxn}) that the $q$-exponential function 
\begin{equation} 
e_q^{\alpha x} = \sum_{n=0}^{\infty} \frac{\alpha^nx^n}{[n]_q!},  
\end{equation}
for any complex number $\alpha$, becomes an eigenfunction of $a$, and the normalized coherent states 
of the $q$-oscillator are given by  
\begin{equation} 
\psi_{\alpha,q}(x) = \frac{1}{\sqrt{e_q^{|\alpha|^2}}}\,e_q^{\alpha x} 
                          = \frac{1}{\sqrt{e_q^{|\alpha|^2}}}\,\sum_{n=0}^{\infty} \frac{\alpha^n}{\sqrt{[n]_q!}}\,\xi_{n,q}(x).    
\end{equation} 

The development of quantum groups and the representation theory of associated quantum algebras led first to the 
$\left(q^{-1},q\right)$-oscillator \cite{Macfarlane1989, Biedenharn1989} (see also \cite{Sun1989, Hayashi1990}): 
\begin{eqnarray} 
\left[N,a^\dagger\right] & = & a^\dagger,  \quad  \left[N,a\right] = -a,  \quad  \left[a,a^\dagger\right] = \phi(N+1)-\phi(N),  \nonumber \\ 
\phi(N) & = & \frac{q^{-N}-q^N}{q^{-1}-q} = [N]_{\left(q^{-1},q\right)}.  
\label{qqoscalg}
\end{eqnarray} 
With the relation 
\begin{equation}
[N+1]_{\left(q^{-1},q\right)}-q^{-1}[N]_{\left(q^{-1},q\right)} = q^N, 
\end{equation} 
the relation between $a$ and $a^\dagger$ becomes 
\begin{equation} 
aa^\dagger - q^{-1}a^\dagger a = q^N.     
\end{equation} 
Note the $q \longleftrightarrow q^{-1}$ symmetry.  The Fock representation is 
\begin{eqnarray}
a^\dagger|n\rangle & = & \sqrt{[n+1]_{\left(q^{-1},q\right)}}\,|n+1\rangle,  \quad  
a|n\rangle = \sqrt{[n]_{\left(q^{-1},q\right)}}\,|n-1\rangle,  \nonumber \\  
N|n\rangle & = & n|n\rangle,  \quad   n = 0,1,2,\ldots\ ,  
\label{qqFockrep}
\end{eqnarray} 
where 
\begin{equation} 
[n]_{\left(q^{-1},q\right)} = \frac{q^{-n}-q^n}{q^{-1}-q} 
\end{equation} 
defines the ${\left(q^{-1},q\right)}$-basic number.  Note that 
\begin{equation}
[0]_{\left(q^{-1},q\right)} = 0,  \quad   [1]_{\left(q^{-1},q\right)} = 1,  \quad  \lim_{q\longrightarrow 1}[n]_{\left(q^{-1},q\right)} = n.    
\end{equation}  
Now, with the definitions 
\begin{equation}
[0]_{\left(q^{-1},q\right)}! = 1,  \quad  
\left[n\right]_{\left(q^{-1},q\right)}! = \prod_{j=1}^n [j]_{\left(q^{-1},q\right)},  \quad  n = 1,2,\ldots\ , 
\end{equation} 
we have 
\begin{equation} 
|n\rangle = \frac{{a^\dagger}^n}{\sqrt{[n]_{\left(q^{-1},q\right)}!}}\,|0\rangle.
\end{equation}
Defining the ${\left(q^{-1},q\right)}$-exponential function as 
\begin{equation}
e_{\left(q^{-1},q\right)}^x = \sum_{n=0}^{\infty} \frac{x^n}{[n]_{\left(q^{-1},q\right)}!}, 
\end{equation} 
the normalized coherent states of the ${\left(q^{-1},q\right)}$-oscillator (\ref{qqoscalg}) are given by 
\begin{equation}
|\alpha\rangle_{\left(q^{-1},q\right)} 
   = \frac{1}{\sqrt{e_{\left(q^{-1},q\right)}^{|\alpha|^2}}}\,\sum_{n=0}^{\infty} 
      \frac{\alpha^n}{\sqrt{[n]_{\left(q^{-1},q\right)}!}}\,|n\rangle 
   = \frac{1}{\sqrt{e_{\left(q^{-1},q\right)}^{|\alpha|^2}}}\,e_{\left(q^{-1},q\right)}^{\alpha a^\dagger}|0\rangle.  
\end{equation}
The Bargmann-like representation is given by 
\begin{equation}
a^\dagger = x,  \quad  a = D_{\left(q^{-1},q\right)} = \frac{1}{x}\left(\frac{q^{-xD}-q^{xD}}{q^{-1}-q}\right),  \quad  N = xD.  
\end{equation} 
Note that 
\begin{equation} 
D_{\left(q^{-1},q\right)}F(x) = \frac{F\left(q^{-1}x\right)-F(qx)}{\left(q^{-1}-q\right)x},  \quad  
D_{\left(q^{-1},q\right)}x^n = [n]_{\left(q^{-1},q\right)}x^{n-1}.  
\end{equation} 
Then, the monomials 
\begin{equation}
\xi_{n,\left(q^{-1},q\right)}(x) = \frac{x^n}{\sqrt{[n]_{\left(q^{-1},q\right)}!}},  \quad  n = 0,1,2,\ldots\ ,  
\end{equation}  
carry the Fock representation of the $\left(q^{-1},q\right)$-oscillator algebra (\ref{qqoscalg}) as given by 
\begin{eqnarray}
a^\dagger\xi_{n,\left(q^{-1},q\right)} & = & \sqrt{[n+1]_{\left(q^{-1},q\right)}}\,\xi_{n+1,\left(q^{-1},q\right)},  \nonumber \\ 
a\xi_{n,\left(q^{-1},q\right)} & = & \sqrt{[n]_{\left(q^{-1},q\right)}}\,\xi_{n-1,\left(q^{-1},q\right)},  \nonumber \\ 
N\xi_{n,\left(q^{-1},q\right)} & = & n\xi_{n,\left(q^{-1},q\right)},  \quad  n = 0,1,2,\ldots\ . 
\end{eqnarray} 
The monomials 
$\{\xi_{n,\left(q^{-1},q\right)}(x)|n = 0,1,2,\ldots \}$ form an orthonormal basis with respect to the inner product 
\begin{equation}
\langle f|g\rangle_{\left(q^{-1},q\right)} = \left. \left[f^*\left(D_{\left(q^{-1},q\right)}\right)g(x)\right]\right|_{x=0}.  
\end{equation}
In this Bargmann-like representation, the $\left(q^{-1},q\right)$-exponential function 
\begin{equation} 
e_{\left(q^{-1},q\right)}^{\alpha x} = \sum_{n=0}^{\infty} \frac{\alpha^nx^n}{[n]_{\left(q^{-1},q\right)}!},  
\end{equation}
for any complex number $\alpha$, is an eigenfunction of $a$, and 
\begin{equation} 
\psi_{\alpha,\left(q^{-1},q\right)}(x) 
   = \frac{1}{\sqrt{e_{\left(q^{-1},q\right)}^{|\alpha|^2}}}\,e_{\left(q^{-1},q\right)}^{\alpha x}   
   = \frac{1}{\sqrt{e_{\left(q^{-1},q\right)}^{|\alpha|^2}}}\,\sum_{n=0}^{\infty}     
      \frac{\alpha^n}{\sqrt{[n]_{\left(q^{-1},q\right)}!}}\,\xi_{n,\left(q^{-1},q\right)}(x) 
\end{equation} 
is a normalized coherent state of the $\left(q^{-1},q\right)$-oscillator (\ref{qqoscalg}).  

Representation theory of two-parameter quantum algebras led to the generalization of the $\left(q^{-1},q\right)$-oscillator to the 
$(p,q)$-oscillator \cite{Chakrabarti1991} (see also \cite{Jannussis1991, Arik1992}):  
\begin{eqnarray}
\left[N,a^\dagger\right] & = & a^\dagger,  \quad  \left[N,a\right] = -a,  \quad 
\left[a,a^\dagger\right] = \phi(N+1)-\phi(N),  \nonumber \\ 
\phi(N) & = & \frac{p^N-q^N}{p-q} = [N]_{(p,q)} .  
\label{pqoscalg}
\end{eqnarray} 
Since 
\begin{equation}
[N+1]_{(p,q)}-p[N]_{(p,q)} = q^N, 
\end{equation} 
we can write 
\begin{equation}
aa^\dagger - pa^\dagger a = q^N.  	
\end{equation} 
Note the $p \longleftrightarrow q$ symmetry.  The Fock representation is 
\begin{eqnarray}
a^\dagger|n\rangle & = & \sqrt{[n+1]_{(p,q)}}\,|n+1\rangle,  \quad  a|n\rangle = \sqrt{[n]_{(p,q)}}\,|n-1\rangle,  \nonumber \\  
N|n\rangle & = & n|n\rangle,  \quad  n = 0,1,2,\ldots\ ,  
\label{pqFockrep}
\end{eqnarray} 
where 
\begin{equation} 
[n]_{(p,q)} = \frac{p^n-q^n}{p-q} 
\label{pqnumber}
\end{equation} 
defines the $(p,q)$-basic number, or the twin-basic number \cite{Jagannathan2006}.  Note that 
\begin{equation}
[0]_{(p,q)} = 0,  \quad  [1]_{(p,q)} = 1,  \quad  \lim_{p,q\longrightarrow 1}[n]_{(p,q)} = n.  
\end{equation}  
Now, with the definitions 
\begin{equation}
[0]_{(p,q)}! = 1,  \quad  \left[n\right]_{(p,q)}! = \prod_{j=1}^n [j]_{(p,q)},  \quad  n = 1,2,\ldots\ , 
\end{equation} 
we have 
\begin{equation} 
|n\rangle = \frac{{a^\dagger}^n}{\sqrt{[n]_{(p,q)}!}}\,|0\rangle.
\end{equation}
Defining the ${(p,q)}$-exponential function as 
\begin{equation}
e_{(p,q)}^x = \sum_{n=0}^{\infty} \frac{x^n}{[n]_{(p,q)}!}, 
\end{equation} 
the normalized coherent states of the ${(p,q)}$-oscillator (\ref{pqoscalg}) are given by 
\begin{equation}
|\alpha\rangle_{(p,q)} = \frac{1}{\sqrt{e_{(p,q)}^{|\alpha|^2}}}\,\sum_{n=0}^{\infty} \frac{\alpha^n}{\sqrt{[n]_{(p,q)}!}}\ |n\rangle 
                                = \frac{1}{\sqrt{e_{(p,q)}^{|\alpha|^2}}}\,e_{(p,q)}^{\alpha a^\dagger}|0\rangle.  
\end{equation} 
The Bargmann-like representation of the $(p,q)$-oscillator algebra (\ref{pqoscalg}) is given by  
\begin{equation}
a^\dagger = x,  \quad a = D_{(p,q)} = \frac{1}{x}\left(\frac{p^{xD}-q^{xD}}{p-q}\right), \quad 
N = xD.  
\end{equation} 
The $(p,q)$-derivative $D_{(p,q)}$ is such that 
\begin{equation}
D_{(p,q)}F(x) = \frac{F(px)-F(qx)}{(p-q)x},  \quad D_{(p,q)}x^n = [n]_{(p,q)}x^{n-1}.  
\label{Dpq}
\end{equation}
The monomials 
\begin{equation}
\xi_{n,(p,q)}(x) = \frac{x^n}{\sqrt{[n]_{(p,q)}!}}, \quad  n = 0,1,2,\ldots\ ,  
\end{equation}  
carry the Fock representation:  
\begin{eqnarray}
a^\dagger\xi_{n,(p,q)} & = & \sqrt{[n+1]_{(p,q)}}\,\xi_{n+1,(p,q)},  \quad  
a\xi_{n,(p,q)} = \sqrt{[n]_{(p,q)}}\,\xi_{n-1,(p,q)},  \nonumber \\   
N\xi_{n,(p,q)} & = & n\xi_{n,(p,q)},  \quad  n = 0,1,2,\ldots\ . 
\end{eqnarray}
The monomials $\{\xi_{n,(p,q)}(x)|n=0,1,2,\ldots\}$ form an orthonormal basis with respect to the inner product 
\begin{equation}
\langle f|g\rangle_{(p,q)} = \left. \left[f^*\left(D_{(p,q)}\right)g(x)\right]\right|_{x=0}.  
\end{equation}
Note that, in this Bargmann-like representation, the $(p,q)$-exponential function 
\begin{equation} 
e_{(p,q)}^{\alpha x} = \sum_{n=0}^{\infty} \frac{\alpha^nx^n}{[n]_{(p,q)}!},  
\end{equation}
for any complex number $\alpha$, is an eigenfunction of $a$ and a normalized coherent state of the $(p,q)$-oscillator is given by 
\begin{equation} 
\psi_{\alpha,(p,q)}(x) = \frac{1}{\sqrt{e_{(p,q)}^{|\alpha|^2}}}\,e_{(p,q)}^{\alpha x} 
   = \frac{1}{\sqrt{e_{(p,q)}^{|\alpha|^2}}}\,\sum_{n=0}^{\infty} \frac{\alpha^n}{\sqrt{[n]_{(p,q)}!}}\,\xi_{n,(p,q)}(x).  
\end{equation} 

Note that the canonical boson oscillator (\ref{bosealg}), the $q$-oscillator (\ref{qoscalg}), and the $\left(q^{-1},q\right)$-oscillator 
(\ref{qqoscalg}), are special cases of the $(p,q)$-oscillator (\ref{pqoscalg}) corresponding to $(p=1, q=1)$, $p=1$, and $p=q^{-1}$, 
respectively.  The canonical fermion oscillator corresponds to $(p = -1, q = 1)$.  The $q$-fermion oscillator (\cite{Parthasarathy1991, 
Chaichian1990}) and the Tamm-Dancoff oscillator (\cite{Odaka1991, Chaturvedi1993}) are also special cases of the $(p,q)$-oscillator 
corresponding to $p = -q^{-1}$ and $p = q$, respectively.  Actually, one can choose $p$ as any function of $q$ in building models.  Thus, 
starting with the  $(p,q)$-oscillator a plethora of interesting types of $q$-deformed oscillators have been found with their energy spectra 
having accidental pairwise energy level degeneracies \cite{Gavrilik2007, Gavrilik2008a, Gavrilik2008b}.  Further, there exist several other 
deformed oscillator structures with single and multiple deformation parameters (see, {\em e.g.}, \cite{Katriel1996, Burban2007, 
Hounkonnou2007, Gavrilik2010, Chung2020}).  It may also be noted that $(p,q)$-calculus based on the $(p,q)$-number (\ref{pqnumber}) 
and the $(p,q)$-derivative (\ref{Dpq}), sometimes called the post-quantum calculus following the name quantum calculus given to the 
calculus based on the $q$-number (\ref{qno}) and the $q$-derivative (\ref{qD}) (see \cite{Kac2002}), is a very active area of research in 
applied mathematics with applications to approximation theory, computer-aided geometric design, etc. (see, {\em e.g.}, \cite{Mursaleen2016, 
Alotaibi2018, Soontharanon2020} and references therein).  

It is known since the early days of deformed oscillators that the creation and annihilation operators of the deformed oscillator (\ref{phialg}) 
can be realized in terms of the canonical boson creation and annihilation operators as 
\begin{equation}
a^\dagger = \sqrt{\frac{\phi(N)}{N}}b^\dagger, \quad 
a = b\sqrt{\frac{\phi(N)}{N}} = \sqrt{\frac{\phi(N+1)}{N+1}}b, \quad N = b^\dagger b.   
\end{equation}
Writing $\phi(N) = f^2(N)N$, the deformed oscillator (\ref{phialg}) has been presented as  
\begin{eqnarray}
a^\dagger & = & f(N)b^\dagger, \quad a = bf(N) = f(N+1)b, \quad N = b^\dagger b,  \nonumber \\ 
\left[N,a^\dagger\right] & = & a^\dagger, \quad [N,a] = -a,  \nonumber \\ 
\left[a,a^\dagger\right] & = & f^2(N+1)(N+1)-f^2(N)N,    
\label{fosc}
\end{eqnarray}  
called an $f$-oscillator, and a general theory of $f$-oscillators has been developed \cite{Manko1997} (see also \cite{Shanta1994, Sunilkumar2000}) with many applications to nonlinear physics including nonlinear coherent states, or $f$-coherent states, relevant for quantum optics 
(see, {\em e.g.}, \cite{Dudinets2017}; see also \cite{Manko1993a, Manko1993b, Manko1995}).  The function $f(n)$, called the nonlinearity function, characterizes the  $f$-oscillator (\ref{fosc}).  In terms of $f(n)$ the deformed exponential function (\ref{ephix}) can be rewritten as 
\begin{equation}
e_\phi^x = \sum_{n=0}^{\infty}\frac{x^n}{f^2(n)!n!} = e_f^x,     
\label{efx}
\end{equation} 
where 
\begin{equation} 
f^2(0)! = 1, \quad f^2(n)! = \prod_{j=1}^n f^2(j),\ \ \mbox{for}\ n \geq 1. 
\end{equation} 
Then, the normalized coherent states of the $f$-oscillator (\ref{fosc}), eigenstates of $a = f(N+1)b$, are given by 
\begin{equation}
|\alpha\rangle_f = \frac{1}{\sqrt{e_f^{|\alpha|^2}}}\,\sum_{n=0}^{\infty}\frac{\alpha^n}{f(n)!\sqrt{n!}}\,|n\rangle,    
\label{fcoherent}
\end{equation} 
where $f(0)! = 1$, $f(n)! = \prod_{j=1}^n f(j)$ for $n \geq 1$, and $\alpha$ is such that $e_f^{|\alpha|^2} < \infty$.  These states 
(\ref{fcoherent}) are called nonlinear coherent states, or $f$-coherent states, of the  boson oscillator and are characterized by the nonlinearity 
function $f(n)$.  They correspond to the nonlinear coherent states of photons in the context of quantum optics.  When $f(n) = 1/\sqrt{n}$ the 
$f$-coherent state (\ref{fcoherent}) becomes   
\begin{equation}
|\alpha\rangle_h = \sqrt{1-|\alpha|^2}\sum_{n=0}^{\infty}\alpha^n|n\rangle,  \quad  \mbox{where}\ \ |\alpha| < 1, 
\label{hstate}
\end{equation} 
known variously as phase coherent state \cite{Lerner1970}, harmonious state \cite{Sudarshan1993}, or pseudothermal state \cite{Dodonov1995}  (see, {\em e.g.}, \cite{Dodonov2002} for more details).  

\section{Tsallis $q$-exponential function as a $\phi$-exponential function}  
It has been shown in \cite{Borges1998} that expansion of the Tsallis $q$-exponential function in Taylor series leads to the expression  
\begin{eqnarray}
e_q(x) & = & 1+\sum_{n=1}^{\infty}\frac{Q_{n-1}}{n!}x^n,  \nonumber \\ 
Q_0 & = & 1,\quad Q_n = q(2q-1)(3q-2)\ldots(nq-(n-1)),  \quad  n = 1,2,\ldots\ . 
\label{Borges} 
\end{eqnarray} 
Writing the expression for $e_f^x$ in (\ref{efx}) as 
\begin{equation} 
e_f^x = 1+\sum_{n=1}^{\infty}\frac{x^n}{f^2(n)!n!}, 
\end{equation}
and comparing this expression with (\ref{Borges}) one identifies $e_q(x)$ with $e_f^x$ corresponding to 
\begin{equation}
f(0)! = 1, \quad f(n)! = \frac{1}{\sqrt{Q_{n-1}}}, \quad  n = 1,2,\ldots\ .   
\end{equation}
In \cite{Bendjeffal2019} this choice of $f(n)!$ has been substituted in (\ref{fcoherent}) to get the normalized states 
\begin{eqnarray}
|\alpha\rangle 
   & = & \mathcal{N}(\alpha)\left\{|0\rangle + \sum_{n=1}^{\infty}\sqrt{Q_{n-1}}\ \frac{\alpha^n}{\sqrt{n!}}|n\rangle\right\},  
            \nonumber \\  
\mathcal{N}(\alpha) & = & \frac{1}{\sqrt{1+\sum_{n=1}^{\infty}Q_{n-1}\frac{|\alpha|^2}{n!}}}, 
\label{Bendjeffal}
\end{eqnarray}  
called the $f$-coherent states attached to the Tsallis $q$-exponential function.  This class of $f$-coherent states and their nonclassical 
properties and other physical aspects have been studied in detail for $1 < q \leq 2$ in \cite{Bendjeffal2019}.  

Noting that in (\ref{Borges}) 
\begin{equation} 
Q_{n-1} = (1+(q-1)(n-1))! =  \prod_{j=1}^n (1+(q-1)(j-1)),  \quad  n = 1,2,,\ldots\ ,      
\end{equation}
the Taylor series expression for $e_q(x)$ has been identified in \cite{Jaganathan2005} with a deformed exponential function as 
\begin{equation}
e_q(x) = \sum_{n=0}^{\infty}\frac{x^n}{[n]_{(q-1)}!},  
\label{eqx} 
\end{equation}
where  
\begin{eqnarray}
[n]_{(q-1)} & = & \frac{n}{1+(q-1)(n-1)},\quad  n = 0,1,2,\ldots\ ,  \nonumber \\ 
{}[0]_{q-1}! & = & 1,\ \ \left[n\right]_{(q-1)}! = \prod_{j=1}^n [j]_{(q-1)} = \frac{n!}{(1+(q-1)(n-1))!},\ \mbox{for}\ n \geq 1.  \nonumber \\ 
   &   &  
\end{eqnarray} 
Note that 
\begin{equation} 
[0]_{(q-1)} = 0, \quad [1]_{(q-1)} = 1, \quad \lim_{q\longrightarrow 1}[n]_{(q-1)} = n.  
\end{equation}
Regarding $[n]_{(q-1)}$ as a deformed number, a scheme of $q$-deformation of nonlinear maps was proposed in \cite{Jaganathan2005} 
and this proposal has been found to have many applications (see, {\em e.g.}, \cite{Patidar2009, Patidar2011, Banerjee2011, Behnia2017, 
Balakrishnan2017, Canovas2019}).  Equation (\ref{eqx}) shows that we can write $e_q(x)$ as a $\phi$-exponential function:  
\begin{equation}
e_q(x) = \sum_{n=0}^{\infty} \frac{x^n}{\phi_T(n)!} = e_{\phi_T}^x,\ \ \mbox{with}\ \ \phi_T(n) = [n]_{(q-1)}.   
\label{eqxphiT}
\end{equation} 

Let us close this section with an interesting observation.  Note that we can write 
\begin{equation}
e_q(x) = e^{\ln\left((1+(1-q)x)^{\frac{1}{1-q}}\right)} = e^{\sum_{n=1}^{\infty}\frac{(q-1)^{n-1}}{n}x^n}, 
\label{eqx2exp}
\end{equation} 
showing explicitly $\lim_{q\longrightarrow 1}e_q(x) = e^x$.  The expression (\ref{eqx2exp}) for  $e_q(x)$ as an ordinary exponential could 
be useful for approximations in applications.  In particular, for the Tsallis $q$-Gaussian function we can write, for $q \approx 1$,  
\begin{equation}
e_q\left(-\beta x^2\right) \approx e^{-\beta x^2+\frac{q-1}{2}\beta^2x^4-\frac{(q-1)^2}{3}\beta^3x^6}.  
\end{equation} 
In \cite{Pourahmadi1984} (see also \cite{Sachkov1996}) it has been shown that if 
\begin{equation}
h(x) = \sum_{n=1}^{\infty}a_nx^n, 
\end{equation} 
then 
\begin{equation}
e^{h(x)} = \sum_{n=0}^{\infty} c_nx^n,  
\end{equation} 
where $c_0 = 1$, $c_1 = a_1$, and 
\begin{equation}
c_n = a_n+\frac{1}{n}\sum_{j=1}^{n-1}jc_{n-j}a_j, \quad  n = 2,3,\ldots\ .   
\label{crecurrence}
\end{equation} 
Using this result the $q$-exponential function $e_q^x$ in (\ref{qexp}) has been expressed as the exponential of an infinite series in 
\cite{Quesne2004}.  For the Tsallis $q$-exponential function we have 
\begin{eqnarray}
h(x) & = & \sum_{n=1}^{\infty}\frac{(q-1)^{n-1}}{n}x^n,  \nonumber \\  
e^{h(x)} & = & e_q(x) = \sum_{n=0}^{\infty}\frac{x^n}{[n]_{(q-1)}!} = 1+x+\sum_{n=2}^{\infty}\frac{(1+(q-1)(n-1))!}{n!}x^n.  
                       \nonumber \\ 
   &   &  
\end{eqnarray} 
We find $c_0 = 1$ and $c_1 = a_1 = 1$ as should be.  Now, substituting 
\begin{equation}
a_n = \frac{(q-1)^{n-1}}{n}, \quad 
c_n = \frac{(1+(q-1)(n-1))!}{n!}, \quad \mbox{for}\ n \geq 2,   
\end{equation}  
in (\ref{crecurrence}) we get an identity: for $n \geq 2$, 
\begin{eqnarray} 
   &   & \frac{(1+(n-1)(q-1))!}{n!} \nonumber  \\ 
   &   & \ \quad = \frac{(q-1)^{n-1}}{n}+\frac{1}{n}\sum_{j=1}^{n-1}\frac{(q-1)^{j-1}(1+(n-j-1)(q-1))!}{(n-j)!}.  
\label{identity}
\end{eqnarray}  
We can verify this identity directly.  Let us take $(q-1) = 1/\tau$.  Then the identity (\ref{identity}) becomes 
\begin{equation}
(\tau)_n = (n-1)!\tau\sum_{j=0}^{n-1}\frac{(\tau)_j}{j!}, 
\label{alphaid}
\end{equation} 
where $(\tau)_n = \tau(\tau+1)(\tau+2)\ldots(\tau+(n-1))$ is the rising factorial, or the Pochhammer symbol, with 
$(\tau)_0 = 1$.  Now, observe that 
\begin{equation}
(\tau)_{n+1} = n!\tau\sum_{j=0}^{n}\frac{(\tau)_j}{j!}   
               = n\left[(n-1)!\tau\left(\frac{(\tau)_n}{n!}+\sum_{j=0}^{n-1}\frac{(\tau)_j}{j!}\right)\right] 
               = (\tau+n)(\tau)_n,  
\end{equation}  
showing that $(\tau)_n$ given by the right hand side of (\ref{alphaid}) satisfies the defining recurrence relation for $(\tau)_n$, namely, 
$(\tau)_{n+1} = (\tau+n)(\tau)_n$ with $(\tau)_0 = 1$.   

As observed in \cite{Chung2019b}, we can represent the Tsallis $q$-exponential function as a hypergeometric series (see, {\em e.g.}, \cite{Andrews1999} for details of the theory of hypergeometric series) as follows: 
\begin{equation} 
e_q(x) = {}_1F_0\left(\frac{1}{q-1};-;(q-1)x\right),   
\end{equation}  
where 
\begin{equation} 
{}_1F_0(a;-;z) = \sum_{n=0}^\infty (a)_n\frac{z^n}{n!}. 
\end{equation} 
This relation can be verified as follows:  
\begin{eqnarray} 
e_q(x) & = & \sum_{n=0}^\infty\ \frac{x^n}{[n]_{q-1}!} 
             =  1 + \sum_{n=1}^\infty\ \left(\prod_{j=1}^n (1+(q-1)(j-1))\right)\frac{x^n}{n!}   \nonumber \\ 
          & = & 1 + \sum_{n=1}^\infty\ \left(\prod_{j=1}^n \ \left(\frac{1}{q-1}+(j-1)\right)\right)\frac{((q-1) x)^n}{n!}   \nonumber \\ 
          & = & \sum_{n=0}^\infty\ \left(\frac{1}{q-1}\right)_n\frac{((q-1)x)^n}{n!}  \nonumber \\ 
          & = & {}_1F_0\left(\frac{1}{q-1};-;(q-1)x\right).  
\end{eqnarray} 
One may also identify $e_q(x)$ with ${}_2F_1\left(\frac{1}{q-1},b;b;(q-1)x\right)$, where $b$ is an arbitrary nonzero parameter, since  
${}_1F_0(a;-;x) = {}_2F_1(a,b;b;x)$, where  
\begin{equation} 
{}_2F_1(a,b;c;z) = \sum_{n=0}^\infty \frac{(a)_n(b)_n}{(c)_n}\frac{z^n}{n!}  
\end{equation} 
is the Gauss hypergeometric function.  

For the Tsallis $q$-logarithm we get  
\begin{equation} 
\ln_q(1+x) = x\,{}_2F_1(q,1;2;-x).  
\label{qlog2F1} 
\end{equation}   
The proof of this relation is as follows:  
\begin{eqnarray} 
\ln_q(1+x) & = & \frac{(1+x)^{1-q}-1}{1-q} 
                   = \frac{1}{1-q}\left(\sum_{n=0}^\infty\ \left(\begin{array}{c} 
                                                                                   1-q \\ n 
                                                                                   \end{array}\right)x^n-1\right)  \nonumber \\ 
                & = & x\sum_{n=0}^\infty\ (q)_n\frac{(-x)^n}{(n+1)!} 
                   =    x\sum_{n=0}^\infty\ \frac{(q)_n(1)_n}{(2)_n}\frac{(-x)^n}{n!}  \nonumber \\ 
                & = & x\,{}_2F_1(q,1;2;-x).  
\end{eqnarray} 
Using the well known Euler integral representation of ${}_2F_1(a,b;c;z)$ it can be verified that 
\begin{equation} 
 x\,{}_2F_1(q,1;2;-x) = \int_0^1 dt\ \frac{x}{(1+xt)^q} = \frac{(1+x)^{1-q}-1}{1-q} = \ln_q(1+x).  
\end{equation} 
Note that when $q \longrightarrow 1$ the relation (\ref{qlog2F1}) becomes the well known relation $\ln(1+x) = x\,{}_2F_1(1,1;2;-x)$.   
We hope these identifications would help further analysis and generalizations of the Tsallis $q$-exponential and $q$-logarithm functions.  

\section{The deformed oscillator and the deformed derivative associated with the Tsallis $q$-exponential function}  
Let us now consider the deformed oscillator with the commutation relations 
\begin{eqnarray}
\left[N,a^\dagger\right] & = & a^\dagger, \quad \left[N,a\right] = -a, \quad 
\left[a,a^\dagger\right] = \phi_T(N+1)-\phi_T(N), \nonumber \\ 
\phi_T(N) & = & [N]_{(q-1)} = \frac{N}{1+(q-1)(N-1)}.      
\label{Tsallisqalg} 
\end{eqnarray} 
The energy spectrum of this deformed oscillator is given by 
\begin{eqnarray}
E_{n,T} & = & \frac{1}{2}\left(\phi_T(n+1)+\phi_T(n)\right)  \nonumber \\ 
            & = & \frac{1}{2}\left(\frac{2(q-1)n^2+2n+2-q}{(q-1)^2n^2+(3-q)(q-1)n+2-q}\right),  \nonumber \\ 
            &    & \qquad \qquad \qquad \qquad \qquad \qquad \qquad \qquad   n = 0,1,2,\ldots\ , 
\label{EnT}
\end{eqnarray}
as follows from (\ref{defoscE}).  For $1 \leq q \leq 2$ we find that  
\begin{eqnarray} 
E_{n,T}(q=1) & = & n+\frac{1}{2},  \quad n = 0,1,2,\ldots\ ,  \nonumber \\ 
E_{0,T} & = & \frac{1}{2},\ \ \mbox{for}\ 1 \leq q < 2,  \quad  
E_{0,T} = \frac{1}{2},\ \ \mbox{when}\ q \longrightarrow 2,  \nonumber \\ 
E_{n,T} & = & 1,\ \ \mbox{for all}\ n > 0,\ \ \mbox{when}\ q \longrightarrow 2,  \nonumber \\
\lim_{n \longrightarrow \infty}E_{n,T} & = & \frac{1}{q-1},\ \ \mbox{for}\ 1 \leq q \leq 2,  \nonumber \\ 
E_{n+1,T}-E_{n,T} & = & \frac{2-q}{(q-1)^2n^2+2(q-1)n+q(2-q)} \geq 0,  \nonumber \\ 
                            &    & \qquad \qquad \qquad \qquad \qquad \qquad \mbox{for}\ 1 \leq q \leq 2,  \nonumber \\ 
\lim_{q\longrightarrow 2}\left(E_{n+1,T}-E_{n,T}\right) & = & 0,\ \ \mbox{for all}\ n > 0.      
\label{Toscspectrum}
\end{eqnarray}
From these equations we can conclude as follows.   The deformed oscillator (\ref{Tsallisqalg}) becomes a canonical boson oscillator when 
$q = 1$.   For $1 < q \leq 2$ it has a ground state with energy $E_{0,T} = 1/2$, same as a boson, and an infinity of excited states starting 
with $E_{1,T} = (1/2)+(1/q)$ and increasing in diminishing steps up to the maximum energy $E_{\infty,T} = 1/(q-1) < \infty$. As the value 
of $q$ increases from $1$ to $2$ the width of this band of energies of excited states, $E_{\infty,T}-E_{1,T} = (1/(q(q-1)))-(1/2)$, decreases 
and when $q \longrightarrow 2$ the energy band shrinks to a single energy level with infinite degeneracy ($E_{n,T} = 1$ for all $n > 0$).  
Thus, when $q \longrightarrow 2$ this oscillator becomes a two-level system with a nondegenerate ground state with energy $E_{0,T} = 1/2$ 
and an infinitely degenerate excited state with energy eigenvalues $E_{n,T} = 1$ for all $n > 0$.  It should be interesting to investigate the 
physical applications of such deformed oscillators related to the Tsallis $q$-exponential function.  The deformed oscillator algebra 
(\ref{Tsallisqalg}) has been derived in \cite{Chung2019c} in a completely different context, unrelated to the Tsallis $q$-exponential function, 
and the thermodynamical properties of a gas of photons obeying such an algebra have been evaluated (see also \cite{Chung2019a} for a 
discussion of the algebra (\ref{Tsallisqalg})).   

It is seen that the $\mu$-oscillator, introduced in \cite{Jannussis1993},  for which $\phi(N) = N/(1+\mu N)$, with $\mu > 0$, is only slightly 
different from (\ref{Tsallisqalg}).  The $\mu$-oscillator is the first deformed oscillator found with its energies lying within a band of finite width 
exactly like in the case of the deformed oscillator (\ref{Tsallisqalg}).  For $\mu = 0$ the $\mu$-oscillator becomes a canonical boson oscillator 
and for any $\mu > 0$ it has a spectrum bounded from above, but it does not share the other interesting properties of the deformed oscillator (\ref{Tsallisqalg}).  This can be seen by comparing (\ref{Toscspectrum}) with the energy spectrum of the $\mu$-oscillator:   
\begin{eqnarray}
E_{n,\mu} & = & \frac{1}{2}\left(\frac{n+1}{1+\mu(n+1)}+\frac{n}{1+\mu n}\right)   \nonumber \\ 
                & = & \frac{1}{2}\left(\frac{2\mu n^2+2(\mu +1)n+1}{\mu^2n^2+2\mu n+\mu^2+\mu +1}\right), \nonumber \\  
E_{n,\mu=0}  & = & n+\frac{1}{2},  \quad n = 0,1,2,\ldots\ ,  \nonumber \\  
E_{0,\mu} & = &  \frac{1}{2\left(\mu^2+\mu +1\right)},\ \ \mbox{for}\ \mu \geq 0,  \nonumber \\   
\lim_{n \longrightarrow \infty}E_{n,\mu} & = & \frac{1}{\mu},\ \ \mbox{for}\ \mu > 0,  \nonumber \\ 
E_{n+1,\mu}-E_{n,\mu} & = & \frac{1}{\mu^2n^2+2\mu(\mu +1)n+2\mu +1},\ \ \mbox{for}\ \mu \geq 0.  
\end{eqnarray}    
In particular, note that there is no finite value of $\mu$ for which $E_{n,\mu}$ becomes independent of $n$ for all $n > 0$.  It is this property 
which makes the deformed oscillator (\ref{Tsallisqalg}) become a two-level system when $q \longrightarrow 2$. 

The deformed oscillator (\ref{Tsallisqalg}) can be presented as an $f$-oscillator as follows:    
\begin{eqnarray} 
a^\dagger 
   & = & f_T(N)b^\dagger, \quad a = bf_T(N) = f_T(N+1)b, \quad N = b^\dagger b,  \nonumber \\ 
f_T(N) 
   & = & \sqrt{\frac{\phi_T(N)}{N}} = \sqrt{\frac{[N]_{(q-1)}}{N}} 
      =   \frac{1}{\sqrt{(1+(q-1)(N-1)))}}.  
\end{eqnarray} 
It has been found in \cite{Bendjeffal2019} that the states in (\ref{Bendjeffal}) are the coherent states of this $f$-oscillator corresponding to the eigenfunctions of $a = f_T(N+1)b$ and are related to the Tsallis $q$-exponential function.  To understand this let us proceed as follows.  The 
Fock representation of (\ref{Tsallisqalg}) is given by 
\begin{eqnarray}
a^\dagger|n\rangle & = & \sqrt{[n+1]_{(q-1)}}\ |n+1\rangle,  \quad   a|n\rangle = \sqrt{[n]_{(q-1)}}\ |n-1\rangle,  \nonumber \\  
N|n\rangle & = & n|n\rangle, \quad  n = 0,1,2,\ldots\ .   
\end{eqnarray} 
It follows that 
\begin{equation} 
|n\rangle = \frac{{a^\dagger}^n}{\sqrt{[n]_{(q-1)}!}}|0\rangle.
\end{equation}
With the Tsallis $q$-exponential function expressed as in (\ref{eqx}), the normalized coherent states of the deformed oscillator 
(\ref{Tsallisqalg}) are seen to be given by 
\begin{equation}
|\alpha\rangle_T = \frac{1}{\sqrt{e_q\left(|\alpha|^2\right)}}\sum_{n=0}^{\infty} \frac{\alpha^n}{\sqrt{[n]_{(q-1)}!}}|n\rangle 
                        = e_q\left(\alpha a^\dagger\right)|0\rangle.  
\label{nlcs}
\end{equation}
Noting that 
\begin{equation}
|n\rangle = \frac{{a^\dagger}^n}{\sqrt{[n]_{(q-1)}!}}|0\rangle 
             = \frac{\left(f_T(N)b^\dagger\right)^n}{\sqrt{[n]_{(q-1)}!}}|0\rangle 
             = \frac{f_T(n)!{b^\dagger}^n}{\sqrt{[n]_{(q-1)}!}}|0\rangle,  
\end{equation}
and using (\ref{efx}), (\ref{fcoherent}), and (\ref{eqxphiT}), it is seen that 
\begin{equation}  
|\alpha\rangle_T = \frac{1}{\sqrt{e_q\left(|\alpha|^2\right)}}\sum_{n=0}^{\infty} \frac{f_T(n)!\sqrt{n!}\alpha^n}{[n]_{(q-1)}!}|n\rangle   
   = \frac{1}{\sqrt{e_{f_T}^{|\alpha|^2}}}\sum_{n=0}^{\infty} \frac{\alpha^n}{f_T(n)!\sqrt{n!}}|n\rangle,  
\end{equation} 
is an $f$-coherent state of the boson oscillator associated with the Tsallis $q$-exponential function as identified in \cite{Bendjeffal2019}.  

To get the Bargmann-like representation of the deformed oscillator (\ref{Tsallisqalg}) we must have a deformed derivative $D_{(T,q)}$ such 
that 
\begin{equation}
D_{(T,q)}x^n = [n]_{(q-1)}x^{n-1}.  
\label{TsallisDq}
\end{equation} 
Following (\ref{phiBargmann}), as has been done in (\cite{Chung2019c, Chakrabarti2010, Kim2019}), one may choose the formal operator 
$(1/x)\phi(xD) = (1/(1+(q-1)xD))D$ as the required deformed derivative which satisfies (\ref{TsallisDq}) and has $e_q(kx)$ as its 
eigenfunction for any $k$.  However, in this case it is only a formal operator, not suitable to operate on an arbitrary function without a series expansion.  Another suggestion for the required deformed derivative is $(1+(1-q)x)D$ for which $e_q(x)$ is an eigenfunction (see 
\cite{Borges2004, Chung2019d}).  But, it is also not suitable for the purpose since $e_q(kx)$ is not its eigenfunction for arbitrary $k$ 
unless one defines $e_q(kx) = (e_q(x))^k$ as in \cite{Carlitz1956, Carlitz1979, Chung2019b}.  As we have seen earlier, for the Tsallis 
$q$-exponential function $e_q(kx) = (1+(1-q)kx)^{1/(1-q)} \neq e_q(x)^k$.   Further, the derivative  $(1+(1-q)x)D$ does not satisfy the 
requirement in (\ref{TsallisDq}).  

To derive the required deformed derivative we follow \cite{Rebesh2013} where the deformed derivative corresponding to the $\mu$-oscillator \cite{Jannussis1993} has been obtained.  Slightly modifying the suggestion in \cite{Rebesh2013} we 
get the desired deformed derivative as 
\begin{equation}
D_{(T,q)}F(x) = \int_{0}^{1}dt\ t^{1-q}DF\left(t^{q-1}x\right).  
\label{TDq}
\end{equation}
It can be verified that for $F(x) = x^n$ 
\begin{eqnarray}
D_{(T,q)}x^n & = & \int_{0}^{1}dt\ t^{1-q}D\left(t^{(q-1)n}x^n\right) = \int_{0}^{1}dt\ t^{(q-1)(n-1)}Dx^n   \nonumber \\               
                    & = & nx^{n-1}\int_{0}^{1}dt\ t^{(q-1)(n-1)} = \frac{nx^{n-1}}{1+(q-1)(n-1)} \nonumber \\ 
                    & = & [n]_{(q-1)}x^{n-1},  
\end{eqnarray} 
as required.  Then, it follows that 
\begin{equation}
D_{(T,q)}e_q(kx) = D_{(T,q)}\left(\sum_{n=0}^{\infty}\frac{k^nx^n}{[n]_{(q-1)}!}\right) 
                 = ke_q(kx).  
\end{equation} 
It can also be verified directly that 
\begin{equation}
D_{(T,q)}e_q(kx) = ke_q(kx).  
\label{DTqalpha}
\end{equation} 
This is seen as follows.  From (\ref{TDq}) we have 
\begin{eqnarray}  
D_{(T,q)}e_q(kx) & = & \int_{0}^{1}dt\ t^{1-q}D\left[\left(1+(1-q)kt^{q-1}x\right)^{\frac{1}{1-q}}\right]  \nonumber \\ 
    & = & k\int_{0}^{1}dt\ \left(1+(1-q)kt^{q-1}x\right)^{\frac{q}{1-q}}  \nonumber \\ 
    & = & k\int_{0}^{1}dt\ t^{-q}\left(t^{1-q}+(1-q)kx\right)^{\frac{q}{1-q}}  \nonumber \\    
    & = & \left.\left[\frac{k}{1-q}\int d(t^{1-q})\ \left(t^{1-q}+(1-q)kx\right)^{\frac{q}{1-q}}\right]\right|_{t=0}^{t=1}  \nonumber \\     
    & = & \left.\left[kt\left(1+(1-q)kt^{q-1}x\right)^{\frac{1}{1-q}}\right]\right|_{t=0}^{t=1} \nonumber \\
    & = & \left.\left[kte_q\left(kt^{q-1}x\right)\right]\right|_{t=0}^{t=1} = ke_q(kx),  
\label{DTqeqkx} 
\end{eqnarray} 
proving (\ref{DTqalpha}).  The inverse of the deformed derivative $D_{(T,q)}$, the deformed integral, is seen to be given by 
\begin{equation}
\int d_{(T,q)}x\ f(x) = \left.\left[\frac{d}{dt}\left(\int dx\ tf\left(t^{(q-1)}x\right)\right)\right]\right|^{t=1}_{t=0},  
\label{TIntq} 
\end{equation} 
such that if $\int d_{(T,q)}x\ f(x) = F(x)$ then $D_{T,q}F(x) = f(x)$.  It can be verified that 
\begin{equation}
\int d_{(T,q)} x\ x^n = \left.\left[\frac{d}{dt}\left(\int dx\ t^{(q-1)n+1}x^n\right)\right]\right|^{t=1}_{t=0} 
                              = \frac{x^{n+1}}{[n+1]_{(q-1)}}, 
\end{equation} 
as should be, since $D_{(T,q)}x^{n+1} = [n+1]_{(q-1)}x^n$.  

Now, the Bargmann-like representation of the deformed oscillator (\ref{Tsallisqalg}) is given by 
\begin{equation}
a^\dagger = x, \quad a= D_{(T,q)}, \quad N = xD.  
\end{equation} 
Under the inner product 
\begin{equation}
\langle f|g\rangle_T = \left. \left[f^*\left(D_{(T,q)}\right)g(x)\right]\right|_{x=0}, 
\end{equation} 
the monomials 
\begin{equation}
\xi_{n,T}(x) = \frac{x^n}{\sqrt{[n]_{(q-1)}!}}, \quad  n = 0,1,2,\ldots\ , 
\end{equation} 
form an orthonormal basis and carry the Fock representation  
\begin{eqnarray}
a^\dagger\xi_{n,T}(x) & = & \sqrt{[n+1]_{(q-1)}}\,\xi_{n+1,T}(x),  \quad  
a\xi_{n,T}(x) = \sqrt{[n]_{(q-1)}}\,\xi_{n-1,T}(x), \nonumber \\ 
N\xi_{n,T}(x) & = & n\xi_{n,T}(x), \quad  n = 0,1,2,\ldots\ .   
\end{eqnarray}
In the Bargmann-like representation, with 
\begin{equation}
\left\langle e_q(\alpha x)\right.\left|e_q(\alpha x)\right\rangle_T = e_q\left(|\alpha|^2\right), 
\end{equation} 
the normalized coherent states of the deformed oscillator (\ref{Tsallisqalg}), eigenfunctions of $a = D_{(T,q)}$, are given by 
\begin{eqnarray}
\psi_{\alpha,T}(x) & = & \frac{1}{\sqrt{e_q\left(|\alpha|^2\right)}}\,e_q(\alpha x)
      =  \frac{1}{\sqrt{e_q\left(|\alpha|^2\right)}}\,\sum_{n=0}^{\infty}\frac{\alpha^nx^n}{[n]_{(q-1)}!}  \nonumber \\   
   & = & \frac{1}{\sqrt{e_q\left(|\alpha|^2\right)}}\,\sum_{n=0}^{\infty}\frac{\alpha^n}{\sqrt{[n]_{(q-1)}!}}\,\xi_{n,T}(x).     
\end{eqnarray}

The case $q = 2$ is interesting.  In this case we have $f_T(n) = 1/\sqrt{n}$ and the state $|\alpha\rangle_T$ becomes the harmonious state (\ref{hstate}) as observed in \cite{Bendjeffal2019}.  Now, for $q = 2$ we have 
\begin{equation}
[0]_{(1)} = 0,  \quad  [n]_{(1)} = 1,\ \  \mbox{for all}\ n.  
\end{equation} 
Thus, for $q = 2$ the deformed oscillator (\ref{Tsallisqalg}) has the Fock representation    
\begin{eqnarray}
a^\dagger|n\rangle & = & |n+1\rangle,  \quad  n = 0,1,2,\ldots\ ,  \nonumber \\ 
a|0\rangle & = & 0,  \quad  a|n\rangle = |n-1\rangle,  \quad  n = 1,2,\ldots\ , \nonumber \\ 
N|n\rangle & = & n|n\rangle,  \quad  n = 0,1,2,\ldots\ , 
\end{eqnarray} 
and can be identified with the algebra of the Susskind-Glogower exponential phase operators (see,  {\em e.g.}, \cite{Manko1997}).  The 
coherent states of this deformed oscillator are 
\begin{equation}
|\alpha\rangle_{(T,q=2)} = \sqrt{1-|\alpha|^2}\sum_{n=0}^{\infty}\alpha^n|n\rangle, 
                         \quad  |\alpha| < 1, 
\end{equation} 
where $\sqrt{1-|\alpha|^2} = 1/\sqrt{e_2\left(|\alpha|^2\right)}$ in which  $e_2\left(|\alpha|^2\right)$ is a Tsallis $q$-exponential function corresponding to $q = 2$.  

It is instructive to look at the deformed derivative and integral operators explicitly for $q = 2$.  Corresponding to $q = 2$ the deformed 
derivative becomes, as seen from (\ref{TDq}), 
\begin{equation}
D_{(T,q=2)}F(x) = \int_{0}^{1}dt\ t^{-1}DF(tx).
\end{equation} 
For $F(x) = x^n$ 
\begin{equation} 
D_{(T,q=2)}x^n = \int_{0}^{1}dt\ nt^{n-1}x^{n-1} = x^{n-1},   
\end{equation} 
as required.  To verify that 
\begin{equation} 
e_2(kx) = \frac{1}{1-kx}, 
\end{equation} 
is an eigenfunction of $D_{(T,q=2)}$, satisfying  
\begin{equation}
D_{(T,q=2)}e_2(kx) = ke_2(kx),     
\end{equation} 
we follow (\ref{DTqeqkx}) to get 
\begin{eqnarray}  
D_{(T,q=2)}e_2(kx) 
   & = & \int_{0}^{1}dt\ t^{-1}D\left[(1-ktx)^{-1}\right] \nonumber \\ 
   & = & k\int_{0}^{1}dt\ (1-ktx)^{-2}  \nonumber \\ 
   & = & k\int_{0}^{1}dt\ t^{-2}\left(t^{-1}-kx\right)^{-2}  \nonumber \\ 
   & = & \left.\left[-k\int d(t^{-1})\ \left(t^{-1}-kx\right)^{-2}\right]\right|_{t=0}^{t=1}  \nonumber \\     
   & = & \left.\left[kt\left(1-ktx\right)^{-1}\right]\right|_{t=0}^{t=1} \nonumber \\ 
   & = & \left.\left[kte_2(ktx)\right]\right|_{t=0}^{t=1} = ke_2(kx).  
\end{eqnarray}

In this case the deformed integral is, as seen from (\ref{TIntq}), 
\begin{equation}
\int d_{(T,q=2)}x\ f(x) 
   = \left.\left[\frac{d}{dt}\left(\int dx\ tf(tx)\right)\right]\right|^{t=1}_{t=0}. 
\end{equation} 
For $f(x) = x^n$ 
\begin{equation}
\int d_{(T,q=2)}x\ x^n = \left.\left[\frac{d}{dt}
                 \left(\int dx\ t^{n+1}x^n\right)\right]\right|^{t=1}_{t=0} = x^{n+1},  
\end{equation} 
as should be, since $D_{(T,q=2)}x^{n+1} = x^n$.   

\section{Conclusion} 
In fine, it is found that the deformed oscillator defined by the commutation relations 
\begin{eqnarray}
\left[N,a^\dagger\right] & = & a^\dagger, \quad \left[N,a\right] = -a,  \nonumber \\ 
\left[a,a^\dagger\right] & = & \phi_T(N+1)-\phi_T(N),  \nonumber \\ 
                                  &    & \qquad \mbox{with}\ \ \phi_T(N) = \frac{N}{1+(q-1)(N-1)}, 	
\end{eqnarray} 
is associated with the Tsallis $q$-exponential function 
\begin{equation} 
e_q(x) = (1+(1-q)x)^{\frac{1}{1-q}},  
\end{equation} 
representable as 
\begin{equation} 
e_q(x) = \sum_{n=0}^{\infty}\frac{x^n}{\phi_T(n)!},\ \ \mbox{with}\ \phi_T(n)! = \phi_T(n)\phi_T(n-1)\ldots\phi_T(2)\phi_T(1).      
\end{equation}
In a Bargmann-like representation the annihilation operator $a$ corresponds to a deformed derivative defined by  
\begin{equation} 
D_{(T,q)}F(x) = \int_{0}^{1}dt\ t^{1-q}DF\left(t^{q-1}x\right), 
\label{Tderive}
\end{equation} 
with the Tsallis $q$-exponential functions as its eigenfunctions.  Thus the Tsallis $q$-exponential functions are the coherent states of the 
deformed oscillator (\ref{Tsallisqalg}).  When $q = 2$ these deformed oscillator coherent states correspond to states known variously as phase coherent states, harmonious states, or pseudothermal states.  Further, when $q = 1$ this deformed oscillator is a canonical boson oscillator, 
when $1 < q < 2$ it has an energy spectrum bounded from above and with the ground state same as a boson, and when $q \longrightarrow 2$ 
it becomes a two-level system with a nondegenerate ground state and an infinitely degenerate excited state.  It should be worthwhile to study 
the physical applications of such deformed oscillators.  The expression in  (\ref{Tderive}) for the deformed derivative $D_{(T,q)}$ for which the Tsallis $q$-exponential functions are eigenfunctions should lead to interesting applications, particularly, in nonextensive statistical mechanics which is 
essentially based on the Tsallis $q$-exponential function and its inverse function, namely, the Tsallis $q$-logarithm function.  A remark in \cite{Rebesh2013}, made in the context of the $\mu$-oscillator, points to the existence of possibilities for extending the deformed derivative 
(\ref{Tderive}) and the Tsallis $q$-exponential function with more deformation parameters.  For example, following the remark in 
\cite{Rebesh2013}, if we substitute the ordinary derivative $D$ in (\ref{Tderive}) by the two-parameter derivative $D_{(p,r)}$ ($D_{(p,q)}$ in (\ref{Dpq}) with $q$ replaced by $r$) we would get a $(p,q,r)$-deformed derivative and its eigenfunctions would be the Tsallis 
$(p,q,r)$-exponential functions (see \cite{Schwammlea2007} for the existing two-parameter generalization of the Tsallis $q$-logarithm and 
$q$-exponential function).

\end{document}